# FORECAST FOR THE NEXT EON: APPLIED COSMOLOGY AND THE LONG-TERM FATE OF INTELLIGENT BEINGS[1]

**Milan M. Ćirković**

*Astronomical Observatory Belgrade*

*Volgina 7, 11160 Belgrade, SERBIA*

*e-mail:* `arioch@eunet.yu`

# FORECAST FOR THE NEXT EON: APPLIED COSMOLOGY AND THE LONG-TERM FATE OF INTELLIGENT BEINGS

**Abstract.** Cosmology seems extremely remote from everyday human practice and experience. It is usually taken for granted that cosmological data cannot rationally influence our beliefs about the fate of humanity—and possible other intelligent species—except perhaps in the extremely distant future, when the question of heat death (in an ever-expanding universe) becomes actual. Here, an attempt is made to show that it may become a practical issue much sooner, if an intelligent community wishes to maximize its creative potential. New developments in the fields of anthropic self-selection and physical eschatology give solid foundations to such a conclusion. This may open some new (and possibly urgent) issues in the areas of future policy making and transhumanist studies generally. It may also give us a slightly better perspective on the SETI endeavor.

> The end of our foundation is the knowledge of causes, and secret motions of things; and the enlarging of the bounds of human empire, to the effecting of all things possible.
>
> Francis Bacon, *The New Atlantis* (1626)

## 1. INTRODUCTION: PHYSICAL ESCHATOLOGY

Living beings change their environments and are changed by their environments in turn. This simple truth has become especially pertinent within the framework of astrobiology. Even before the onset of the explosive development of this field we are witnesses of, the fact that even simple lifeforms can influence its physical and chemical environment on the planetary scale has been widely known. The stock example is the one of the Earth's atmosphere, which is markedly out of chemical equilibrium due to the presence of the biosphere, and has been so for billions of years. As the author of the Gaia hypothesis, James Lovelock wrote: "Almost everything about its composition seems to violate the laws of chemistry... The air we breathe... can only be an artifact maintained in a steady state far from chemical equilibrium by biological properties." (Lovelock 1988) Recently publicized projects dealing with detection of exoplanetary biospheres all rely on this simple fact. In somewhat different light, we are all sadly aware of the impact of human activities on the biosphere, Earth's climate and Earth's circumplanetary space.

It is reasonable, at least, to conclude that the magnitude and spatial extent of these biological influences increase with time spanned by the activities of the life forms considered. In that case, the interplay between physical and biological evolution of any chosen environment has to be disentangled with special care when characteristic timescales of the two types of processes become similar. These difficulties and uncertainties are well-illustrated by recent fierce debates over the causes of the short-term climatic change. Since it deals with the very widest conceivable environments, the question of relevant timescales for changes induced by life and intelligence is very relevant to the astrobiological endeavor. Human existence so far is too short for the conclusions in this respect to be drawn from experience. In order to consider the impact of humans (and, by analogy, other intelligent communities), on the surrounding universe, we need to investigate what physics may tell us about its future.

Physical eschatology is a rather young branch of astrophysics, dealing with the future fate of astrophysical objects, as well as the universe itself. Landmark studies in physical eschatology are those of Rees (1969), Dyson (1979), Tipler (1986) and Adams and Laughlin (1997). Some relevant issues have been discussed in the monograph of Barrow and Tipler (1986), as well as several popular-level books (Islam 1983; Davies 1994; Adams and Laughlin 1999); detailed bibliography can be found in Ćirković (2002). Since the distinction between knowledge in classical cosmology and physical eschatology depends on the distinction between past and future, several issues in the physics and philosophy of time are relevant to the assessment of eschatological results and *vice versa*.

As of recently, we have been witnessing a true revolution in cosmology, promising for the first time since Einstein and de Sitter, to finally fix the "Big Three" basic cosmological parameters: cosmological matter density $\Omega$, Hubble constant $H_0$, and cosmological constant $\Lambda$ (for a classical textbook treatment of these see Peebles 1993). Tremendous success of the two projects of observing supernovae of Type Ia in distant galaxies (Perlmutter et al. 1999; Riess et al. 2001), as well as constantly improving observations of the microwave background radiation (Lineweaver 1998; Knox, Christensen, and Skordis 2001) are most to be credited for this revolution. These observational projects have demonstrated that the expansion of the universe is, contrary to our intuitions, not decellerating, but on the contrary: universe is expanding at ever-increasing rate! This is possible only if there is some sort of "exotic" stuff dominating the dynamics and acting as "antigravity"; since Einstein, we know for such a candidate: cosmological constant. According to the emerging picture, the universe is globally flat—corresponding to $\Omega = 1$—but is dominated by cosmological constant (vacuum energy) which makes for about 70% of the total energy density. (In technical terms, $\Omega = \Omega_m + \Omega_\Lambda$, $\Omega_m$ being the contribution of all matter, and $\Omega_\Lambda$ the one of the cosmological constant; we now have $\Omega_\Lambda \approx 0.7$, $\Omega_m \approx 0.3$.) The present expansion is exponentially accelerating from a turning point, a few billion years ago, when the dynamical effects of the cosmological constant first overwhelmed those of gravitating matter.

The outlook this paradigm suggests for the future is rather bleak. Universe will not only expand forever, but will do so at ever-increasing speed. Gravitationally bound structures, such as galaxies, galaxy clusters and superclusters will become more and more isolated in the future. As noticed by several investigators (Rindler 1956; Krauss and Starkman 2000; Ćirković and Bostrom 2001), cosmological constant acts to create an *event horizon*, i.e. closed surface across which communication is impossible *at all times*. This effectively means that any perturbation larger the size of horizon cannot affect us. On the other hand, the inexorable rise of entropy will degrade matter configurations in each of the regions forever surrounded by event horizons, and the state very close to the classically imagined heat-death (e.g. Eddington 1931) will ensue. The processes of star-formation and stellar nucleosynthesis, at present the major sources of entropy production, will cease, and the remaining stellar remnants will be slowly degraded by proton decay and gravitational collapse. Even black holes will inevitably evaporate on tremendously long timescales through the Hawking evaporation. Finally, nothing will remain except the incredibly redshifted thermal photons of wavelengths comparable to the horizon size, since even the remaining not annihilated electrons and positrons will be separated by distances far surpassing the horizon size.

A necessary ingredient in most serious discussions of physical eschatology is the presence of living and intelligent systems in the future of the universe (which *ex hypothesi* did not exist in its past). Dyson has been the first to boldly spell it out in 1979:

> It is impossible to calculate in detail the long-range future of the universe without including the effects of life and intelligence. It is impossible to calculate the capabilities of life and intelligence without touching, at least peripherally, philosophical questions. If we are to examine how intelligent life may be able to guide the physical development of the universe for its own purposes, we cannot altogether avoid considering what the values and purposes of intelligent life may be. But as soon as we mention the words value and purpose, we run into one of the most firmly entrenched taboos of twentieth-century science.

The future of universes containing life and intelligence is *essentially* different from the future of universes devoid of such forms of complex organized matter; as well as different from the past of the same universes in which complexity was lower. This is the crucial point of contact with the astrobiological discourse. In a similar vein, John A. Wheeler has written in a beautiful paper on the relationship of quantum mechanics and cosmology (Wheeler 1988):

> Minuscule though the part is today that such acts of observer-participancy play in the scheme of things, there are billions of years to come. There are billions upon billions of living places yet to be inhabited. The coming explosion of life opens the door to an all-encompassing role for observer-participancy: to build, in time to come, no minor part of what we call *its* past—*our* past, present and future—but this whole vast world.

Obviously, most of the discussions of place and the long-term future of intelligent observers in the universe rely on some assumptions pertaining to the relevant motivations of intelligent communities. Various assumptions have been used in the existing literature, the most interesting one being that the expansion of such communities and the consequent technologization of space are carried out by particular technical means, notably von Neumann probes (von Neumann 1966; Tipler 1986, 1994). However, even its most fervent supporters do not claim that such a course of action on the part of intelligent communities is necessary, exclusive, or even dominant. On the contrary, the strength of the celebrated Fermi's "paradox" lies in its necessity of exclusiveness of extraterrestrial behaviour (e.g. Brin 1983; Hanson 1998a,b). To various arguments invoked to support the conjecture that the expansion and colonization of space are generic characteristics of intelligent communities, one may add the one which we shall attempt to describe in this essay, formulated as the natural generalization of the concept of self-interest. The plan of exposition is as follows. Before we consider the main task, in Sections 3 and 4, it is necessary to define an extremely useful auxiliary notion (Section 2). Sec. 5 summarizes our ideas and a couple of rather technical issues of cosmology are relegated to Appendices, for the sake of easier reading.

## 2. OBSERVER-MOMENTS AND SELF-SAMPLING ASSUMPTION

When examining the possibility of life and intelligence playing a significant role on ever-larger spatial and temporal scales, one essential constraint to take into account is the so-called Doomsday Argument (henceforth DA; for a survey of already voluminous literature on the subject, see Leslie 1996; Bostrom 2001a, 2002; Olum 2002). Roughly, DA argues from our temporal position, according to a principle that directly corresponds to the way other applications of anthropic reasoning argue from the expected typicality of our position in the universe. It was conceived (but not published) by the astrophysicist Brandon Carter in the early 1980s, and it has been first exposed in print by John Leslie in 1989 and in a *Nature* article by Richard Gott in 1993. The most comprehensive discussion of the issues involved is Leslie's monograph of 1996, *The End of The World*. The core idea can be expressed through the following urn-ball experiment. Place two large urns in front of you, one of which you know contains ten balls, the other a million, but you do not know which is which. The balls in each urn are numbered 1, 2, 3, 4, ... Now take one ball at random from the left urn; it shows the number 7. This clearly is a strong indication that the left urn contains only ten balls. If the odds originally were fifty-fifty (identically-looking urns), an application of Bayes' theorem gives the posterior probability that the left urn is the one with only ten balls as $P_{post} (n=10) = 0.99999$. Now consider the case where instead of two urns you have two possible models of humanity, and instead of balls you have human individuals, ranked according to birth order. One model suggests that the human race will soon become extinct (or at least that the number of individuals will be greatly reduced), and as a consequence the total number of humans that ever will have existed is about 100 billion. The other model indicates that humans will colonize other planets, spread through the Galaxy, and continue to exist for many future millennia; we consequently can take the number of humans in this model to be of the order of, say, $10^{18}$. As a matter of fact, you happen to find that your rank is about sixty billion. According to Carter and Leslie, we should reason in the same way as we did with the urn balls. That you should have a rank of sixty billion is much more likely if only 100 billion humans ever will have lived than if the number was $10^{18}$. Therefore, by Bayes' theorem, you should update your beliefs about mankind's prospects and realize that an impending doomsday is much more probable than you thought previously. Here we are not interested in DA *per se*, but in one notion whose introduction in the field of anthropic thinking has been motivated by DA.

Namely, DA and similar probabilistic arguments have been grounded in the basic equality of all observers within a reference class.[i] However, this may be insufficient in most realistic situations, and may well misrepresent the actual contribution of the attribute "intelligent" to the ontological status of an "intelligent observer". Therefore, Bostrom (2002) makes the following attempt at definition, which we shall accept in further discussion:

> ...We can take a first step towards specifying the sampling density by substituting "*observer-moments*" for "observers". Different observers may live differently long lives, be awake different amounts of time, spend different amounts of time engaging in anthropic reasoning etc. If we chop up the stretch of time an observer exists into discrete observer-moments then we have a natural way of weighing in these differences. We can redefine the reference class to consist of all observer-moments that will ever have existed. That is, we can upgrade SSA to something we can call the *Strong Self-Sampling Assumption*:

> *(SSSA)* Every observer at every moment should reason as if their present observer-moment were randomly sampled from the set of all observer-moments.

An additional motivation for introducing observer-moments comes directly from thinking about the future: it is difficult to predict the properties of future observers, in particular their longevity and the metabolic/information-processing rates. For instance, the DA conclusion may turn out to be perfectly correct if humanity achieves immortality coupled with zero population growth; it seem obviously unfair to count observers (instead of observer-moments) in the same manner before and after transition to the "immortal" regime. Thus, counting observer-moments may be a much more tractable approach, since one may absorb all changes in, say, metabolic rate—via Dyson's biological scaling hypothesis (Dyson 1979), or convenient generalizations—in the simple arithmetic changes in the budget of observer-moments. This also offers the simplest unifying framework for treating various kinds of observers, originating at various locations in spacetime. As we shall now see, however, the tally of observer-moments is influenced by cosmological factors in two different ways. The first and most obvious one is contained in the relevant limits that follow from cosmological boundary conditions. The second, dealing with the impact of cosmological studies on the possible social and technological policies of intelligent communities, however, has not been treated in the literature so far.

Following SSSA, we obtain a method of quantitatively comparing the measure of "success" of different (actual or possible) civilizations. Plausibly, one may expect that an advanced civilization will seek to maximize its total tally of observer-moments, since it is a necessary precondition for maximization of any other existential or creative goals. Let us denote this conjecture with a special symbol:

> **Conjecture (\*):** An intelligent community tends to maximize its total number of observer-moments, *ceteris paribus*.

Of course, we are assuming here a long-term behaviour (since we are interested in investigation of interaction between cosmological and intelligent factors). Obviously, the validity of the conjecture (\*) is highly uncertain, since we know so little about the nature (and especially physical groundings) of intelligence and mental phenomena in general. However, we wish to argue here that (\*) is useful as a working hypothesis, from which meaningful conclusions may be drawn. Let us note that (\*) is not exclusive: it is enough for our purposes that maximization of the number of observer-moments is just one of the goals of any intelligent society.

Let us denote the total tally of observer-moments of an intelligent community over its entire history by $\Theta$. Thus, the variational form of

$$\delta \Theta = 0 \tag{1}$$

when applied over all possible civilization's *histories* describes the desired future in the most general form, i.e. gives a mathematical expression to our conjecture (\*).[ii] As usual in the variational calculus, the solution is sought which satisfies a number of independent constraints; hypothetical aims an goals of a civilization different from (\*) may be treated as some of such constraints. However, it is illusory to hope to explicate the functional $\Theta$ in such general terms, since the problem is, obviously, extravagantly difficult. Instead, we shall use a greatly simplified "temporal" model, in which we assume that the civilization is characterized by discrete individual observers, countable (with their observer-moments) at any given time. This may be mathematically expressed as:

$$\Theta = \int_{t_{min}}^{t_{max}} N(t)\langle\sigma(t)\rangle\,dt, \qquad (2)$$

where N(t) is the number of observers at epoch t of cosmic time, and $\langle\sigma(t)\rangle$ the corresponding average density of their observer-moments.[iii] The lifetime of the civilization considered spans the interval from $t_{min}$ and $t_{max}$, where the upper limit may—in principle—be infinite. It is important to emphasize that we use physical time here (i.e. we acknowledge the validity of the Weyl postulate which enables one to define a universal, "cosmic" timescale), although it is possible to change coordinates to some subjective timescale if more appropriate, in the manner of Dyson's biological scaling hypothesis (Dyson 1979; Krauss and Starkman 2000), or (in)famous Tipler's Omega-point theory (Tipler 1994).[iv] There are *at least two distinct ways* in which cosmological parameters enter into eq. (2):

1. Most obviously, the values of the cosmological parameters determine absolute limits on $t_{min}$ and $t_{max}$. If the entire lifetime of the universe is equal to $\tau$, then $t_{max} \leq \tau$. In addition, $t_{min} > 0$, but also one may state that $t_{min} \geq \tau_*$, where $\tau_*$ is the epoch of formation of the first stars of sufficiently high metallicity for the processes of chemical and biological evolution to take place.

2. The shape of the function N(t) is dependent on the cosmological parameters when the discrete nature of matter distribution is taken into account. Namely, the power spectrum of cosmological density perturbations determines which objects form as result of gravitational attraction and decoupling from the universal Hubble expansion (for a modern textbook treatment see Peebles 1993). On the other hand, the size of the matter aggregates like stars, galaxies, etc. is essential for answering the question how large a part of the rest mass can be converted into energy for the purpose of (intelligent) information processing. In essence, the cosmological power spectrum is the ultimate source of the famous Kardashev's taxonomy of possible advanced intelligent communities, to which we shall return later.

It is plausible to assume that the *maximal* number of observers is proportional to the energy consumed for such purposes, which can be mathematically written as

$$N_{max}(t) \propto q\int_{V} dV \int_{t_{min}}^{t} dt\,\rho_i, \qquad (3)$$

where $\rho_i$ denotes the relevant energy density, and $q < 1$ is the efficiency of whatever energy extraction process a civilization uses. The reason why we consider the maximal number of observers is that the exact number, of course, depends on the sociological factors which are completely outside the scope of the present study. It may also strongly depend on the level of technology (e.g. Sandberg 2000), and may radically decrease with further scientific and technological advancement (like in the cyberpunk scenarios of "collective consciousness" development). Neglecting this, we perceive that at least this upper limit is still cosmologically determined, since both the relevant densities $\rho_i$ and integration bounds are contained in the cosmological discourse. Of course, the density $\langle \sigma(t) \rangle$ is even less tractable from the point of view of the present knowledge, since it may be expected to hinge crucially upon biological factors of which we know little. Advances in the algorithmic information theory and its application to biological systems, as well as artificial life, might eventually shed some light on this difficult issue (e.g. Chaitin 1977; Langton 1986). However, for the purposes of the present study, it is enough to assume that it is the non-zero function of time which either increases or decreases slower than exponential.

## 3. COSMOLOGICAL REVOLUTION: A STORY

How does the number of observer moments $\Theta$ tally with various cosmological models, including the realistic one? Let us first note that there is doubt as to whether such thing as the *exact* model can ever be reached. Several simplifications come handy at this point. Sufficiently high degree of symmetry leads to familiar Friedmann models (or generalization of them including the cosmological constant), and sufficiently small perturbations can be treated in a familiar way. However, even the general outline on which the future fate of a universe depends may not be obvious until some critical epoch occurs to any internal observers. In particular, as discussed in detail in an illuminating essay by Krauss and Turner (1999), realistic universes are notoriously difficult to analyze completely, due to the possible presence of very large (super-horizon) perturbations which enter the visible universe only at some later epoch. From the point of view of internal observers, there is no possibility to avoid this ambiguity. In such a position, it is natural that the priorities leading to maximization of the number of observer-moments in (2) are contingent on the contemporary cosmological knowledge. As Krauss and Starkman (2000) vividly put it, "funding priorities for cosmological observations will become exponentially more important as time goes on."

Let us now investigate the following imaginary situation. A civilization inhabiting a particular, sufficiently symmetric universe, develops both theoretical and observational astronomy to the point where it can make useful working models of their universe as a whole. After one equivalent of an Einstein of that particular world develops a formalism to describe curved spacetime at the largest scales, an equivalent of Hubble discovers universal expansion, and the equivalents of Penzias and Wilson discover the remnants of primordial fireball, leading cosmologists begin to support the concept of a flat baryonic universe with $\Omega_B = \Omega \approx 1$. At first it seems that all observations can be accomodated in the framework of such a model (we suppose that light elements' abundances, for instant, are not inconsistent with such high baryonic density, contrary to the situation in *our* observable universe!). Some circumstantial support for this model comes from ingenious theoreticians of that civilization, who discover that the coupling of a universal scalar field to gravity leads to exponential expansion during the very early epochs. This inflationary phase in the history of such a universe leads to the prediction that $|\Omega - 1| = \varepsilon \approx 10^{-5}$, while it is not clear whether the universe is marginally closed or marginally open. In the latter case (favored by most of the theoreticians in such a universe), the number of galaxies in their universe is infinite, and therefore such a universe offers a very optimistic prospect for the survival of intelligence and life. There is no event horizon in such a universe, and the particle horizon is (very) roughly given as the age of the universe in light years, i.e. the maximal path traversed by light along the observer's past light cone. What are the prospects of intelligent beings surviving indefinitely in such a universe?

Gradually, bolder scientists begin to tackle physical eschatological issues. An equivalent of Dyson on that world reckons that this civilization can, in principle, indefinitely survive while exploiting its sources of energy in larger and larger volumes ($t_{max} = \infty$). In addition, it was suggested by some extremely speculative and ingenious cosmologists, that a non-zero cosmological shear can be manifested in later epochs, providing in this manner additional energy which will be proportional to the volume of the technologized space (although this option has not been studied enough). The predominant attitude toward the maximization of (2) is, therefore, very optimistic and not characterized by any sense of urgency. There are physical grounds to expect $\Theta_{max} = \infty$. Everybody is happy and relaxed. It seems natural, and even desirable, to seek perfection in domains like the quality of life, individual development and aesthetics, rather than devote energies in projects of space colonization, since there is arguably enough time for everything.

Suddenly, a new and unexpected twist occurs. New cosmological observations, and in particular two superbly designed projects detecting standard candles at large distances in order to make a best-fit estimate of the Hubble constant, indicate a spectacular overthrow of the ruling paradigm. After the dust settles (which lasts for years, and probably decades), the new paradigm suggest that the universe is still geometrically flat, but dominated by the cosmological constant term $\Lambda$ in such a way that $\Omega = \Omega_B + \Omega_\Lambda = 1$, $\Omega_B = 0.1$, $\Omega_\Lambda = 0.9$. Now, the situation radically changes with respect to the envisaged number of possible observer-moments given by (*). The universe is now found to possess not only a particle, but an event horizon also, defined as the surface through which any form of communication is impossible at all epochs. This is a consequence of the fact that after a phase of power-law expansion, the exponential expansion generated by $\Lambda$ sets in, thus creating a second (future and final) inflationary phase in the history of the universe (see Appendix I for some technical details).

There is further bad news for such a civilization. The decrease in the metabolic temperature envisaged by the Dyson-equivalent can not continue indefinitely, as it was possible before the "cosmological revolution", since the de Sitter universe possesses a minimal temperature, a circumstance following from the quantum field theory, and described in some detail in the Appendix I. This is an extremely small temperature, but still finite, and below it nothing can be cooled without expending precious free energy. Thus, the temperature scaling may be continued only to the final value of $t_{max}$ in (2). In addition, one may not use any shear energy, since the equivalent of the so-called "cosmological no-hair" theorem guarantees that no significant shear remains during the exponential expansion (Gibbons and Hawking 1977).

It seems obvious that the "cosmological revolution" will have important social and political consequences if the desire of maximizing $\Theta$ in (2) remains (or becomes) the legitimate goal of the considered civilization. There could be no more leisurely activities in the framework of the second paradigm. Although survival cannot be indefinite, it still seems that it can be prolonged for a very, very long time—but only if one starts early enough. Besides funding for cosmological observations, one may expect that funding for interstellar and even intergalactic expansion will suddenly rise.

Colonization of other stellar and (ultimately) galactic systems should better start early in the Λ-dominated universe!

## 4. DIFFICULTIES INVOLVED IN ESTIMATES

This story can teach us several lessons. It seems that we are currently in the middle of the "cosmological revolution" described above, although a not so dramatic one, since there was never a consensus on the values of cosmological parameters or the nature of matter constituents in the actual human cosmology. Also, as we have mentioned above, the currently inferred value for the vacuum density $\Omega_\Lambda$ is somewhat smaller, being about 0.7 (e.g. Perlmutter et al. 1999; Zehavi and Dekel 1999), corresponding to $\Omega_m = 1 - \Omega_\Lambda \approx 0.3$. However, the qualitative nature of the revolution and the implied potential change in the entire spectrum of human social and technological activities are analogous.

Of course, this counterfactual example may be regarded as rather conservative. One may imagine much more drastic changes in the dominant cosmological paradigm. Let us, for instance, suppose that for some reason most cosmologists did accept the classical steady state theory of Bondi, Gold and Hoyle in late 1940-ies, and that in the same time the development of radio astronomy has been dampened for several more decades. Let us recall the basic tenets of this, now largely forgotten, but historically enormously influential cosmological theory. Bondi and Gold derived the steady state theory from what they called the Perfect Cosmological Principle, which can be simply expressed as the homogeneity of the universe in 4-dimensional spacetime (instead of just in 3-dimensional space, the latter statement being, since Eddington and Milne called the Cosmological Principle). This simple postulate has a tremendous wealth of consequences, the most important (and the most controversial) being the continuous creation of matter at very small rates throughout spacetime. As a consequence, as old galaxies recede toward infinity and disappear (since the steady-state universe also possesses a horizon), new galaxies are formed which take their place in an everlasting and—on large scales—unchanging universe. The same process of creation enables matter to stay at constant average entropy throughout eons and avoid the menace of the heat death.

The attitude of humanitarian thinkers seeking to maximize $\Theta$ could very well be encouraged by the steady state concept of the creation of low-entropy matter in the manner conserving the density of matter fields. Not only did one have $t_{max} = \infty$, one should also expect $\lim_{t\to\infty} N(t) = \infty$, and there would have been no plausible reason to expect $\sigma(t)$ to be anything but constant or even an increasing function of time. From the particular human point of view, therefore, the steady state cosmology offered one of the most optimistic visions of the future.[v] (This is somewhat ironical, since the steady state model predicts essentially the same exponentially expanding spacetime as the Λ-dominated models.) As we know, after the fierce cosmological battle in 1950s and early 1960s, the steady state theory has been finally overthrown by the sensational discoveries of QSOs and the cosmic microwave background, as described in a colorful recent history of Kragh (1996). There has been no historical consensus about the exact cosmological model accounting for observations ever since, but it seems that we are on the verge of reaching one. However, it is conceivable that the cosmology of some other civilization passes directly from the steady state into the Λ-dominated paradigm. This seems, curiously enough, at least in one respect easier and more natural than what has occurred in the actual history (see Appendix II). This paradigm shift *must* be accompanied by a shift in technological and social priorities if one expects $\Theta$ to be maximized. Such implications are present even if our conjecture (*) is correct only selectively or on the average.

However, changes in the cosmological paradigm currently underway in the real world should not be regarded as the end of the story. As mentioned above, perturbations of the scale larger than the horizon scale are expected to enter our visible universe only at some late epoch. In the light of the argument above, one may expect that whatever the cosmological paradigm is established on the timescale of next $\sim 10^1$ years, may be upset by observing the perturbations on superhorizon scales (Krauss and Turner 1999). A recent intriguing study of Tipler (1999) shows that the cosmological conclusions reached by local observations (i.e. those in the vicinity of the Milky Way) can be highly misleading, and that one should be on guard with respect to the results of any local measurement of cosmological parameters.

Let us try to estimate the effects of a belated technologization to the lowest order. It should not be especially emphasized that any such estimate is notoriously difficult, speculative and on the very fringe of the domain of founded scientific hypothesizing; some of the reasons, already mentioned, include our almost perfect ignorance of the evolutionary possibilities in the social domain, as well as the influence of various technological advances on the average census of observer-moments per observer, $\langle \sigma(t) \rangle$. Even the simpler part of the problem, the estimate on the possibilities and modes of evolution of the number of observers $N(t)$, poses almost intractable difficulties. We may be virtually certain that the current exponential population growth of humanity will be arrested at some future date, but whether it will result in transition to some other (power-law?) growing function, or tend to a stable asymptotic limit is impossible to establish at this time. There are certainly several timescales relevant for the history of an advanced technological community, which are related to the "quantized" nature of physical resources alluded to above (and which are, ultimately, consequences of the cosmological power spectrum). This may roughly correspond to Kardashev's famous classification of advanced intelligent communities into three types, depending on the energy resources available (e.g. Tarter 2001 and references therein). By allowing for "ergodicity" we may treat Kardashev's types as temporal phases in (unobstructed) progress of any single intelligent community; we would only like to add, for reasons to be explained below, the "Type IV", corresponding to a civilization successfully managing the energy resources of the largest gravitationally bound structure it originated in. If our position is cosmologically typical, these would be galaxy superclusters. However, there has been no estimates of the timescales required for transition between the types (and possible intermediate timescales corresponding to radically new technologies of energy extraction; for instance the Penrose-Christodoulou process of extracting the gravitational energy of black holes).

The baryonic mass of the Local Supercluster (henceforth LS) is of the order of $10^{15}$ solar masses (Oort 1983, and references therein), and its luminosity several times $10^{12}$ solar luminosities. Let us suppose that humanity will eventually technologize the entire spatial volume of LS, and gather all its negentropy resources for information processing. Let us also suppose that at whatever time humans (or posthumans) embark on the process of galactic and intergalactic colonization, the historical path of such colonization will be essentially the same. This is a reasonable assumption, since we expect that the colonization timescale is significantly smaller from the cosmological timescales characterizing large-scale changes in the distribution of matter within LS. If we further assume (as many of the prominent anthropic thinkers, following Carter's well-known argument, do) that we are the first technological civilization within LS, we may ask the question how many observer-moments (or conceivable human lives and experiences) we loose by postponing the onset of colonization by $\Delta t$? The simplest ("zero-order") estimate is simply to assume that all entropy produced by the physical processes in LS during that interval is proportional to the loss of information from the "pool" available to the presumed "Type IV" future hypercivilization (i.e. the one exploiting the energy resources of LS). The major entropy producing process at present (and on the timescales relevant to the issue; see Adams and Laughlin 1997) is stellar nucleosynthesis. Its products are high-entropy photons escaping to intergalactic (and subsequently intersupercluster) space and being there further redshifted due to the universal expansion. Using the Brillouin (1962) inequality, we may write

$$\Delta I = 5 \times 10^{70} q \left( \frac{L_{SC}}{10^{12} L_s} \right) \left( \frac{\Delta t}{100 \, yrs} \right) \left( \frac{T}{2.73 \, K} \right)^{-1} \text{bits,} \qquad (4)$$

where $L_s$ is the Solar luminosity, and $q$ is the (time-averaged) fraction of free energy which hypercivilization converts into work for its computing devices. We expect that the temperature $T$ at which computations are performed will be close to the temperature of the cosmic microwave background since the timescale even for the colonization of a huge object like LS is short by cosmological standards, and thus such colonization is essentially isothermal. The quantity of information lost per 100 years of delay in starting the colonization is astonishing by any standard. For a conservative estimate of $q = 0.1$, and using Dyson's (1979) estimate of complexity—be it program-size complexity of Chaitin or any similar measure—of an average present-day human being $Q_h \approx 10^{23}$ bits (a quantity which is likely to grow in future, especially in the hypothetical posthuman stage of our evolution, but which is still a useful benchmark), the *number of potentially viable human lifetimes lost* per 100 years of postponing the onset of galactic colonization is simply (if we assume that the luminosity fraction in the equation above is unity, which is probably an underestimate by a factor of a few)

$$\Delta n = \frac{\Delta I}{Q_h} \sim 5 \times 10^{46} . \text{ (!!!)} \qquad (5)$$

Of course, this is only the total integrated loss; if for some currently unknown reason the colonization of LS is impossible or unfeasible, while colonization of some of its substructures is possible and feasible, this huge number should be multiplied by a fraction of the accessible baryonic matter currently undergoing a significant entropy increase (essentially luminous stars). On the other hand, our estimate is actually conservative for the following reasons. There are other entropy-producing processes apart from stellar radiation (notably the stellar black-hole formation which becomes more and more important as the time passes), and therefore the lost quantity of potentially useful information is likely to be higher. Another very pertinent reason why this estimate should be taken as the absolute lower limit is the entire spectrum of *existential risks* (see Bostrom 2001b), which have not been taken into account here.

Namely, the realistic history of posthuman civilization would be the *convolution* of the integrand functions in (2) with a risk function $f_{risk}(t)$ describing the cumulative probability of existential risks up to the epoch $t$ (and their presumed impact on the observer-moment tally):

$$\Theta = \int_{t_{min}}^{t_{max}} f_{risk}(t) N(t) \langle \sigma(t) \rangle \, dt \,. \qquad (6)$$

Obviously, this function would be biased toward higher levels at small values of $t$ (as measured, for instance, from the present epoch for humans), since smaller—i.e. those not colonizing the universe—civilizations are more prone to all sorts of existential risks. In other words, existential risks act to "filter out" observer-moments viable in the long-run from the pool of all conceivable observer-moments associated with a concrete intelligent species.[vi]

Thus, the risk inherent in a "colonization later" policy makes our estimate very conservative (or "optimistic" from the point of view of lost observer-moments). However, this estimate possesses the virtue of being a natural extension of the Dyson concept of development of a Type II (Kardashev) civilization: in order to truly technologize the domicile planetary system, an advanced society must strive to capture and exploit the entire stellar energy output of its home star, via Dyson spheres or similar contraptions (Dyson 1960; Sandberg 2000). *Mutatis mutandis*, the same arguments apply to larger scales of density fluctuations, and in the $\Lambda$-dominated cosmological model we are supplied with a natural cut-off at larger scales.

## 5. SUMMARY

The above said testifies to the simple truth that an awareness of the cosmological situation is a first step toward true long-term planning for any community of intelligent observers interested in self-preservation and achieving maximum creative potential. However, in an evolving universe, the issue of timing seems to set stringent limits on the efficiency with which intelligent communities can fulfil their goals. While those limits are certainly to be the subject of much debate and discussion in the future, the very fact of their existence makes cosmology interesting from the transhumanist perspective. Decision-making performed today, as far as humanity is concerned, may have enormous consequences over very long timescales. In particular, an overly conservative approach to space colonization and technologization, may result (and in fact might have already resulted) in the loss of a substantial fraction of all the possible observer-moments humanity could have had achieved. It is our modest hope that this cursory study will contribute to the wider and livelier discussion of these issues and help to reach other, more precise predictions for intelligence's cosmological future.

Finally, let us note that this approach is not necessarily the only manner in which cosmology may enter our everyday life. If some approaches within the fundaments of quantum mechanics and its links to the human conscience are correct, we may find ourselves in a situation where the cosmological boundary conditions determine the nature of our perceptions and self-awareness (Wheeler 1988; Dugić, Ćirković and Raković 2000). This differs markedly from our approach in this essay, which is based on classical cosmology (as well as classical logic and probability theory). One may imagine that in the future the correct physical theory of conscience will incorporate these elements, and that they will *a fortiori* play some role in any policy-making attempts.

**APPENDIX I**

The behavior of a universe with a large positive vacuum energy density—commonly (and somewhat imprecisely) known as the cosmological constant—$\Lambda$ has been investigated in several recent publications even before the cosmological supernovae began to throw light on its reality (the major review is Carroll, Press and Turner 1992; see also Krauss and Turner 1999; Ćirković and Bostrom 2000). In the $\Lambda$-dominated epoch, the scale factor behaves according to the de Sitter law, i.e.

$$R(t) = R_0 \exp(Ht), \tag{I.1}$$

where the "effective" Hubble constant is given as $H = \sqrt{\Lambda/3}$. The relationship of $\Lambda$ and $\Omega_\Lambda$ is $\Omega_\Lambda = \dfrac{c^2 \Lambda}{3 H_0^2}$, where $c$ is the speed of light, $\Omega_\Lambda$ is the cosmological density of the vacuum and $H_0 \equiv 100 \ h$ km s$^{-1}$ Mpc is the present-day Hubble constant parametrized in such way that $h$ is a dimensionless number of order unity. Obviously, any universe with matter could not immediately find itself in the state of exponential expansion given in (I.1), since at small spatial separations, the effective repulsion due to the cosmological constant is negligible. Thus, in any realistic universe, after a transitional period between matter-domination and vacuum-domination, exponential expansion sets in and the cosmological event horizon (surface impassable to causal influences, similar to event horizons of black holes) is formed. The size of the cosmological event horizon is given as:

$$R_h = \frac{c}{H_0} \Omega_\Lambda^{-0.5} = 3.6 \times 10^9 \ h^{-1} \left( \frac{\Omega_\Lambda}{0.7} \right)^{-0.5} \text{pc}. \tag{I.2}$$

Beyond this distance no communication is possible *at any time*. This is very different from the situation in the matter-dominated universes, where the contribution of the cosmological constant is very small or completely vanishing, where there are only so-called particle horizons, representing temporary obstacles to communication (i.e. any two arbitrarily chosen points will get into a region of causal influence in finite future time).

The minimal temperature of the exponentially expanding (de Sitter) universe characterized by the cosmological constant $\Lambda$ is given by the equation (Gibbons and Hawking 1977):

$$T \to T_\Lambda = \frac{\hbar c}{k} \sqrt{\frac{\Lambda}{12\pi^2}} \approx 3.3 \times 10^{-30} h \sqrt{\frac{\Omega_\Lambda}{0.7}} \text{ K}, \tag{I. 3}$$

where $k$ is the Boltzmann constant. Roughly speaking, this is the temperature of the black-body radiation with maximal emissivity at wavelengths equal to the cosmological event horizon. Modern observations indicate that the expression under the square root on the right-hand side of (I. 2) is close to unity, and h $\approx$ 0.6. Therefore, this temperature is low beyond description, but as longer and longer timescales in the future unfold, its finite value precludes the asymptotic process ($1/T \to \infty$) of the lowering metabolic rate of intelligent creatures of the far future suggested by Dyson (1979) as a method for achieving immortality (Krauss and Starkman 2000).

## APPENDIX II

Ironically enough, it would not be difficult to confuse the classical steady-state cosmology with the Λ-dominated ones if the level of sophistication of (neo)classical cosmological tests (e.g. Sandage 1988) were not very high. Namely, the major *observational* parameter used in empirical discrimination between world models is the *decceleration parameter $q_0$*, defined as

$$q_0 = -\frac{R\ddot{R}}{\dot{R}^2},\qquad\qquad\text{(II.1)}$$

where $R$ is the cosmological scaling factor. Of course, this definition is not of much practical value. Instead, it can be shown that in standard relativistic Friedmann-Robertson-Walker cosmologies, $q_0$ is related to densities in matter and vacuum in the following way (with the usual assumption of negligible pressure):

$$q_0 = \frac{\Omega_m}{2} - \Omega_\Lambda,\qquad\qquad\text{(II.2)}$$

which delivers the "classical" value of 0.5 for the Einstein-de Sitter model ($\Omega = \Omega_m = 1$, $\Omega_\Lambda = 0$), but becomes strongly negative for the vacuum-dominated models. In particular, for the extreme model considered above ($\Omega_m = 0.1$, $\Omega_\Lambda = 0.9$), we have

$$q_0 = -0.85.\qquad\qquad\text{(II.3)}$$

It is well-known that, on the other hand, the decceleration parameter in the steady-state model is

$$q_0 = \text{const.} = -1.\qquad\qquad\text{(II.4)}$$

Obviously, the last two values are close enough for the clear and unequivocal discrimination between them to be an extremely hard observational task.

**Acknowledgements.** I use this opportunity to express my gratitude to Olga Latinović, Vesna Milošević-Zdjelar, Srdjan Samurović, Milan Bogosavljević and Branislav Nikolić for their help in finding some of the references. Further thanks go to Mark Walker and other editors of the *Journal of Evolution and Technology* (http://www.jetpress.org/index.html), who helped significantly improve a previous short draft of the present paper. The manuscript enormously benefited from discussions with Nick Bostrom, Petar Grujić and Fred C. Adams. Kind advice of Robert J. Bradbury, Mark A. Walker and Mašan Bogdanovski is also appreciated.

Technical help of Alan Robertson and of my mother, Danica Ćirković, has been invaluable in concluding this project.

---

[i] The latter presents a separate problem, far from being solved in the field of anthropic thinking. What constitutes a reference class is by no mean clear, and some recent discussions (from different premises!) can be found in Bostrom (2001) and Olum (2001).

[ii] We tacitly assume that $\Theta$ is well defined for each history. This conjecture may be impossible to prove, but it does seem plausible in the light of our belief that the reference class problem *will* eventually be solved.

[iii] The important assumption here is that the histories of intelligent species are *ergodic*, i.e. that the ensemble averaging is the same as temporal averaging. Since ergodicity conjectures are notoriously difficult to prove even for simple physical systems, we cannot hope to improve upon this assumption in the present case. Note, however, that most transhumanist issues are inherently ergodic.

[iv] From the mathematical point of view, such transformation should be non-singular except possibly at the boundary of the relevant region. Such is the case with usually suggested transformations; for instance, in the classical Milne universe, we have the connection between the two timescales as $\tau = t_0 \ln (t/t_0) + t_0$, where $t_0$ is a constant (e.g. Milne 1940). The zero point of t-time occurs in the infinite past of $\tau$-time.

[v] Although, of course, such a future could hardly be called eschatological, since physical eschatology is trivial in an unchanging universe. In addition, there is an entire host of very problematic features of the steady state theory following from the application of the Strong Anthropic Principle, since the very absence of obstacles to unlimited growth of civilizations in such a universe would be the clear sign that there must be a factor sharply limiting their growth—since we have not perceived advanced civilizations of arbitrary age in our past light cone (Tipler 1982; Barrow and Tipler 1986). For the

purposes of our present discussion, however, we are justified in neglecting this complication, since it is always possible to imagine a logically consistent cosmological model that very slowly passes from a quasi-stationary to an evolutionary phase (similar to the historically interesting Eddington-Lemaître model; see Ćirković 2000).

[vi] It is possible that the number of filtered observer-moments is very small in a typical case. This would give rise to the "Great Filter" (cf. Hanson 1998b), explaining the "Great Silence" (cf. Brin 1983) and Fermi's paradox.